\documentclass[twocolumn,prd,showpacs,10pt,superscriptaddress]{revtex4}
\usepackage{graphicx}
\usepackage{amsmath, amssymb}

\def\CC{{\cal C}}

\def\LL{{\cal L}}
\def\MM{{\cal M}}
\newcommand{\KL}{D_{KL}}

\def\tr{{\text{tr}}}

\def\be{\begin{equation}}
\def\ee{\end{equation}}
\def\bea{\begin{eqnarray}}
\def\eea{\end{eqnarray}}

\newcommand{\like}{\LL}

\newcommand{\mdl}{\mathcal{M}}
\newcommand{\lsim}{\,\raise 0.4ex\hbox{$<$}\kern -0.8em\lower 0.62ex\hbox{$\sim$}\,}
\newcommand{\gsim}{\,\raise 0.4ex\hbox{$>$}\kern -0.7em\lower 0.62ex\hbox{$\sim$}\,}

\newcommand{\ie}{{\it i.e.}}
\newcommand{\eg}{{\it e.g.}}
\newcommand{\Otot}{\Omega_{\text{tot}}}

\begin{document}

\title{Measuring the effective complexity of cosmological models}
\date{\today}

\author{Martin Kunz}\email{Martin.Kunz@physics.unige.ch}

\affiliation{D\'{e}partement de Physique Th\'{e}orique,
Universit\'{e} de Gen\`{e}ve, 24 quai Ernest Ansermet, CH-1211
Geneva 4, Switzerland}

\author{Roberto Trotta}\email{rxt@astro.ox.ac.uk}
\affiliation{Oxford University, Astrophysics,  Denys Wilkinson
Building, Keble Road, OX1 3RH, United Kingdom}

\author{David R.~Parkinson}\email{D.R.Parkinson@sussex.ac.uk}

\affiliation{Astronomy Centre,  University of Sussex, Brighton,
BN1 9QH, United Kingdom }

\begin{abstract}
We introduce a statistical measure of the effective model
complexity, called the {\em Bayesian complexity}. We
demonstrate that the Bayesian complexity can be used to
assess how many effective parameters a set of data can
support and that it is a useful complement to the model
likelihood (the evidence) in model selection questions. We
apply this approach to recent measurements of cosmic
microwave background anisotropies combined with the Hubble
Space Telescope measurement of the Hubble parameter. Using
mildly non--informative priors, we show how the 3-year WMAP
data improves on the first-year data by being able to
measure both the spectral index and the reionization epoch
at the same time. We also find that a non-zero curvature is
strongly disfavored. We conclude that although current data
could constrain at least seven effective parameters, only
six of them are required in a scheme based on the
$\Lambda$CDM concordance cosmology. 
\end{abstract}

\maketitle

\section{Introduction}

The quest for a cosmological standard model is being driven by an
increasing amount of high quality observations. An important and
natural question concerns the number of fundamental parameters
of the underlying physical model. How many numbers are necessary
to characterize the Universe? Or in other words, how complex is
the Universe? Generally the term
``complexity'' is employed in a rather loose fashion: a more
complex model is one with a larger number of parameters
that can be adjusted over a large range to fit the model to the
observations. In this paper we will try to measure the effective
number of model parameters that a given set of data can support.
Because of the connection to the data, we will call this the
{\em effective complexity} or {\em Bayesian complexity}. Our
main purpose is to present a statistically sound quantity that
embodies in a quantitative way the above notion of complexity
when a model is compared to the observations.

Bayesian model comparison makes use of an Occam's razor argument
to rank models in term of their quality of fit {\em and} economy
in the number of free parameters. A model with more free
parameters will naturally fit the observations better, but it will
also be penalized for the wasted parameter space that the larger
number of parameters implies. Several studies have made use of
Bayesian model comparison to assess the viability of different
models in the cosmological context
\cite{ev_use,trotta_evidence,nestsamp}. In this work we show that
Bayesian complexity is an ideal complement to model selection in
that it allows to identify the number of effective parameters
supported by the data.

We start by introducing our notation and the fundamentals of
Bayesian statistics and model comparison. We then present the
Bayesian complexity and illustrate its use in the context of a toy
model in section~\ref{sec:linmod}. In section \ref{sec:cosmo} we
apply it to observations of cosmic microwave background
anisotropies and we conclude in section \ref{sec:conclusions}.

\section{Bayesian model selection and complexity}

\subsection{Model comparison}

We first briefly review the basic ingredients of Bayesian
statistics and some relevant aspects of information theory. This
serves both to introduce our notation and to remind the reader of
the main points. We will use a fairly compact notation where
possible and refer the reader to \eg~\cite{MacKay} for the exact
mathematical definitions. Specifically, for an outcome $x$ of a
random variable $X$ we will write $p(x)$ for the probability
distribution function (pdf), ie the probability that $X$ takes a
certain value $x$. In the case of a multi-dimensional parameter
space we will write $p(x)$ as a short form of the joint
probability over all components of $x$, $p(x_1,x_2,\ldots,x_n)$.
The conditional probability of $x$ given $y$ is written $p(x|y)$.

The starting point of our analysis is {\em Bayes theorem}:
 \be p(x|y,I) =
 \frac{p(y|x,I) p(x|I)}{p(y|I)} . \label{eq:bayes}
 \ee
Here, the quantity $I$ represents a collection of all external
hypotheses and our model assumptions.

Given the data $d$ and a model $\MM$ with $n$ free parameters
$\theta$, statistical inference deals with the task of determining
a posterior pdf for the free parameters of the model,
$p(\theta|d,\MM)$. The latter is computed via Bayes theorem as
 \be \label{eq:bayes2}
 p(\theta|d,\MM) = \frac{p(d|\theta,\MM) p(\theta|\MM)}{p(d|\MM)} .
 \ee
On the right--hand side of this equation, $p(d|\theta,\MM)$ is the
probability of obtaining the observed data given the parameter
value $\theta$. Since the observed data are a fixed quantity, we
interpret $p(d|\theta,\MM)$ as a function of $\theta$ and we call
it the {\em likelihood}, $\LL(\theta) \equiv p(d|\theta,\MM)$. The
{\em prior pdf} $p(\theta|\MM)$ embodies our state of knowledge
about the values of the parameters of the model before we see the
data. There are two conceptually different approaches to the
definition of the prior: the first takes it to be the outcome of
previous observations (\ie, the posterior of a previous experiment
becomes the prior for the next), and is useful when updating ones
knowledge about the values of the parameters from subsequent 
observations. For the scope of this paper, it is more appropriate
to interpret the prior as the available parameter space under the
model, which then is collapsed to the posterior after arrival of
the data. Thus the prior constitutes an integral part of our model
specification. In order to avoid cluttering the notation, we will
write $\pi(\theta)=p(\theta|\MM)$, with the model dependence
understood whenever no confusion is likely to arise.

The expression in the denominator of \eqref{eq:bayes2} is a
normalization constant and can be computed by integrating over the
parameters,
 \be \label{eq:evidence}
 p(d|\MM)=\int d\theta\,\pi(\theta)\LL(\theta).
 \ee
This corresponds to the average of the likelihood function under
the prior and it is the fundamental quantity for model comparison.
The quantity $p(d|\MM)$ is called {\em marginal likelihood}
(because the model parameters have been marginalized), or in
recent papers in the cosmological context, the {\em evidence}. In
the following we shall refer to it as to the {\em likelihood for
the model} \footnote{This is an effort to be consistent with common
terminology in the statistical community and to avoid confusion
with the colloquial meaning of the term ``evidence''.}. The
posterior probability of the model is then, using Bayes theorem
again,
 \be p(\MM|d) = \frac{ p(d|\MM) \pi(\MM)}{p(d)},
 \ee
 where $\pi(\MM)$ is the prior for the model.
 The quantity in the denominator on the right--hand side
is just a normalization factor depending on the data alone, which
we can ignore. When comparing two models, $\MM_1$ versus $\MM_2$,
one introduces the {\em Bayes factor} $B_{12}$, defined as the
ratio of the model likelihoods
 \be
  \frac{p(\MM_1|d)}{p(\MM_2|d)} =
  \frac{\pi(\MM_1)}{\pi(\MM_2)}B_{12}
  \ee
In other words, the prior odds of the models are updated by the data
through the Bayes factor. If we do not have any special reason to
prefer one model over the other before we see the data, then
$\pi(\MM_1) = \pi(\MM_2) = 1/2$, and the posterior odds reduce to
the Bayes factor. Alternatively, one can interpret the Bayes
factor as the factor by which our relative belief in the two model
is modified once we have seen the data.

\subsection{A Bayesian measure of complexity }

The usefulness of a Bayesian model selection approach based on the
model likelihood is that it tells us whether the increased
``complexity'' of a model with more parameters is justified by the
data. But the number of free parameters in a model is only the
simplest possible notion of complexity, which we call {\em input
complexity} and denote by $\CC_0$. A much more powerful definition
of model complexity was given by Spiegelhalter et al \cite{dic},
who introduced the {\em Bayesian complexity}, which measures the
number of model parameters that the data can constrain:
 \be \label{eq:C_b_rewritten}
 \CC_b = -2\left(\KL(p,\pi) - \widehat{\KL} \right) .
 \ee
On the right--hand side, $\KL(p, \pi)$ is the Kullback-Leibler
(KL) divergence between the posterior and the prior, representing
the information gain obtained when upgrading the prior to the
posterior via Bayes theorem:
 \be \label{eq:def_KL}
\KL (p,\pi) \equiv  \int p(\theta|d)
\log\frac{p(\theta|d)}{\pi(\theta)} d\theta
 \ee
The KL divergence measures the relative entropy between the two
distributions (e.g.~\cite{MacKay})\footnote{Notice that the KL
divergence is not symmetric (\ie, in general $\KL (P,Q)\neq \KL
(Q,P)$), but it does have the useful properties that $\KL (P,Q)
\geq 0$ (Gibbs inequality), with equality only for $P=Q$, and
that it is invariant under general transformations of the
parameter space.}. In Eq.~\eqref{eq:C_b_rewritten},
$\widehat{\KL}$ is a point--estimate for the KL divergence. If all
parameters are well measured, then the posterior pdf will collapse
into a small region around $\hat{\theta}$, and thus the KL
divergence will approximately be given by $\widehat{\KL} = \log
p(\hat{\theta})/\pi(\hat{\theta})$. By taking the difference, we
compare the effective information gain to the maximum information
gain we can expect under the model, $\widehat{\KL}$. The factor of
2 is chosen so that $\CC_b \rightarrow \CC_0$ for highly
informative data, as shown below. As a point estimator for
$\hat{\theta}$ we employ the posterior mean, denoted
by an overbar. Other choices are possible, and we discuss this
further below.

We can use Eq.~\eqref{eq:def_KL} and Bayes theorem to rewrite
\eqref{eq:C_b_rewritten} as
 \be \label{eq_def_C_b}
 \CC_b =  -2 \int p(\theta|d,\MM) \log \like(\theta) + 2
 \log\like(\hat{\theta}),
 \ee
Defining an effective $\chi^2$ through the likelihood as
$\LL(\theta)\propto \exp(-\chi^2/2)$ (any constant factors drop
out of the difference of the logarithms in
Eq.~\eqref{eq:C_b_rewritten}) we can write the {\em effective
number of parameters as}
  \be \label{eq:complex_as_chi}
  \CC_b =
\overline{\chi^2(\theta)} - \chi^2(\hat{\theta}),
 \ee
where the mean is taken over the posterior pdf. This quantity can
be computed  fairly easily from a Markov chain Monte Carlo (MCMC)
run, which is nowadays widely used to perform the parameter
inference step of the analysis.

We thus see that the effective number of parameters of the model
is not an absolute quantity, but rather {\em a measure of the
constraining power of the data as compared to the predictivity of
the model}, \ie\ the prior. Hence $\CC_b$ depends both on the data
at hand and on the prior available parameter space. In fact, it is
clear that the very notion of ``well measured parameter'' is not
absolute, but it depends on what our expectations are,
\ie\ on the prior. For example, consider a measurement of
$\Otot$, the total energy density of the Universe, expressed in
units of the critical density. The current posterior uncertainty
around $\Otot = 1$ is about $0.02$. Whether this means that we
have ``measured'' the Universe to be flat (\ie\, $\Otot = 1$) or
not depends on the prediction of the model we consider. If we take
a generic prior in the range $0 \leq \Otot \leq 2$, then we
conclude that current data have measured the Universe to be flat
with moderate odds (a precise analysis gives odds of $18:1$ in
favor of the flat model, see \cite{trotta_evidence}). On the
contrary, in the framework of \eg\ landscape theories, the prior
range of the model is much narrower, say $|\Otot -1| \lsim
10^{-5}$, and therefore current posterior knowledge is
insufficient to deem the parameter measured.

Eq.~\eqref{eq:complex_as_chi} is conceptually related to the
$\Delta\chi^2$ approach used for example in \cite{cmpcb} to
analyze how many dark energy parameters are measured by the data.
In that work the decrease of the $\chi^2$ value within the
confidence regions encompassing 68\% and 95\% of the posterior was
measured and compared against the theoretical decrease of a
multivariate Gaussian distribution with a given number of degrees
of freedom (see e.g. chapter 15.6 of \cite{numrep}). This in turn
allowed to deduce the effective number of degrees of freedom of
the $\chi^2$.

We now turn to an explicit demonstration of the power and
usefulness of coupling the model likelihood with the Bayesian
model complexity in model selection questions, by analyzing in
some detail linear toy models.

\section{Linear models \label{sec:linmod}}

Before applying the Bayesian complexity and model likelihood to
current cosmological data, we compute them explicitly for a linear
model and we illustrate their use in a toy example involving
fitting a polynomial of unknown degree. We show below how the
Bayesian complexity tells us how many parameters the data could in
principle constrain given the prior expectations under the model.

\subsection{Specification of the likelihood function \label{sec:linmod1}}

Let us consider the following {\em linear model}
 \be
 y = F \theta + \epsilon
 \ee
where the dependent variable $y$ is a $d$-dimensional vector of
observations, $\theta$ is a vector of dimension $n$ of unknown
regression coefficients and $F$ is a $d\times n$ matrix of known
constants that specify the relation between the input variables
$\theta$ and the dependent variables $y$~\footnote{In the special
case of observations $y_i(x)$ fitted with a linear model $f(x) =
\theta_j X^j(x)$, the matrix $F$ is given by the basis functions
$X^j$ evaluated at the locations $x_i$ of the observations,
$F_{ij} = X^j(x_i)$.}. Furthermore, $\epsilon$ is a
$d$-dimensional vector of random variables with zero mean (the
{\em noise}). If we assume that $\epsilon$ follows a multivariate
Gaussian distribution with uncorrelated covariance matrix $C
\equiv{\rm diag}(\tau_1^2, \tau_2^2, \dots, \tau_d^2)$, then the
likelihood function takes the form
 \be p(y | \theta) =
\frac{1}{(2\pi)^{d/2} \prod_j \tau_j}
\exp\left[-\frac{1}{2}(b-A\theta)^t(b-A\theta)\right],
\label{eq:data_like}
 \ee
where we have defined $A_{ij} = F_{ij}/\tau_i$ and $b_i =
y_i/\tau_i$. This can be cast in the form
 \be
 p(y | \theta) =
 \LL_0
 \exp\left[-\frac{1}{2}(\theta-\theta_0)^t L (\theta-\theta_0)\right],
\label{eq:like}
 \ee
with the likelihood Fisher matrix $L$ given by
 \be
 L \equiv A^t A
 \ee
 and a normalization constant
 \be
 \LL_0 \equiv \frac{1}{(2\pi)^{d/2} \prod_j \tau_j}
 \exp\left[-\frac{1}{2}(b - \theta_0 A)^t(b-A \theta_0) \right].
 \ee
Here $\theta_0$ denotes the parameter value that maximizes the
likelihood, given by
 \be
 \theta_0 = L^{-1}A^t b.
 \ee
As a shortcut, we will write
 \be \label{eq:defchisq}
 \chi^2(\theta) \equiv -2 \ln p(y|\theta) = \chi^2(\theta_0) + (\theta-\theta_0)^t
 L
 (\theta-\theta_0),
 \ee
 where $\chi^2(\theta_0) = -2 \ln \LL_0$.

 \subsection{Model likelihood and complexity}

In this section we gain some intuitive feeling about the
functional dependence of the model likelihood and complexity on
the prior and posterior for the simple case of linear models
outlined above. The results are then applied in section
\ref{sec:toymodel} to an explicit toy model, showing the model
likelihood and complexity in action.

Assuming as a prior pdf a multinormal Gaussian distribution with
zero mean and Fisher information matrix $P$ (we remind the reader
that the Fisher information matrix is the inverse of the
covariance matrix), \ie
 \be \label{eq:gaussprior}
 \pi(\theta) = \frac{|P|^{1/2}}{(2\pi)^{n/2}}\exp\left[-\frac{1}{2}
 \theta^t P \theta \right] ,
 \ee
where $|P|$ denotes the determinant of the matrix $P$, the model
likelihood \eqref{eq:evidence} and model complexity
\eqref{eq:complex_as_chi} of the linear model above are given by
Eqs.~\eqref{eq:modlike} and \eqref{eq:dic_gauss}, respectively
(see Appendix \ref{ap:computed}).

Let us now consider the explicit illustration of a model with $n$
parameters, $\theta=(\theta_1,\dots, \theta_n)$ and $\CC_0=n$.
Without losing generality, we can always choose the units so that
the prior Fisher matrix is the unity matrix, \ie\
 \be
 P = \text{Id}_n.
 \ee
This choice of units is natural since it is the prior that sets
the scale of the problem. The likelihood Fisher matrix being a
symmetric and positive matrix, it is characterized by $n(n+1)/2$
real numbers, which we choose to be its eigenvalues
$1/\sigma_{i}^{2}$, $i=1,\dots,n$ and the elements of the
orthogonal matrix $U$ that diagonalizes it (corresponding to
rotation angles). Here the $\sigma_{i}$ represent the standard
deviations of the likelihood covariance matrix along its
eigendirections $i$, expressed in units of the prior width. With
$D=\text{diag}(\sigma_1^{-2}, \dots, \sigma_n^{-2})$ we have that
the likelihood Fisher matrix is given by $L=U D U^t$ and thus
Eq.~\eqref{eq:dic_gauss} gives
  \be \label{eq:complex2}
 \begin{aligned}
 \CC_b & = \CC_0 - \text{tr}
 \left[ \left( UDU^t + \text{Id}_n\right)^{-1}\right] =   \\
 & = \CC_0 - \text{tr}\left[ \left( D+ \text{Id}_n \right)^{-1}\right]
   = \sum_{i=1}^n\frac{1}{1+\sigma_i^2}.
\end{aligned}
 \ee
The complexity only depends on how well we have measured the
eigendirections of the likelihood function {\em with respect to
the prior}. Every well--measured direction, (\ie, one for which
$\sigma_i \ll 1$) counts for one parameter in $\CC_b$, while
directions whose posterior is dominated by the prior, $\sigma_i
\gg 1$, do not count towards the effective complexity. This
automatically takes into account strong degeneracies among
parameters. Notice also that once an eigendirection is well
measured (\ie\ in the limit $\sigma_i \ll 1$), then the prior
width does not matter anymore.

In contrast, the model likelihood \eqref{eq:modlike} is given by
(assuming for simplicity that the mean of the likelihood
corresponds to the prior mean, \ie\ $\theta_0 = 0$)
 \be
   p(d | \mdl)  = \LL_0 \prod_{i=1}^n
\frac{\sigma_i}{\sqrt{1+\sigma_i^2}} \label{eq:evid_nd}.
 \ee

Finally, we remark that an important ingredient of the Bayesian
complexity is the point estimator for the KL divergence. Here 
we adopt the posterior mean as an estimate, but other simple
alternatives are certainly possible, for instance the posterior
peak (or mode), or the posterior median. The choice of an optimal estimator
is still matter of research (see \eg~section \ref{sec:cosmo} and the 
comments at the end of
Ref.~\cite{dic}). The important aspect is that the posterior pdf
should be summarized by only one number, namely the value plugged
into the KL estimator. This is obviously going to be a very bad
description for highly complex pdf's, exhibiting for instance
long, banana--shaped degeneracies. No single number can be
expected to summarize accurately such a pdf. On the other hand,
for fairly Gaussian pdf's all the different estimators (mean,
median and peak) reduce to the same quantity. This clearly calls
for using normal directions in parameter space~\cite{kosowsky},
which make the posterior as Gaussian as possible, a procedure
that it would be wise to follow whenever possible for many other
good reasons (\eg, better and faster MCMC convergence).

\subsection{Effective complexity as a data diagnostics}

We now turn to the question of how we can use the model likelihood
and complexity together as a tool for model selection. We shall
see that the Bayesian complexity provides a way to assess the
constraining power of the data with respect to the model at hand
and to break the degeneracy among models with approximately equal
model likelihood.

Let us consider two models $A$ and $B$ with different numbers of
parameters, $n_B > n_A$ (but in general the two models need not to
be nested). If the additional parameters of model $B$ are required
by the data, then the likelihood of model $B$ will be larger and
$B$ will have larger posterior odds, thus it should be ranked
higher in our preference than model $A$. However, even if the
extra parameters of model $B$ are not strictly necessary, they can
lead to over--fitting of the data and compensate the Occam's
penalty factor in (\ref{eq:evid_nd}) sufficiently to lead to a
comparable marginal likelihood for both models.  The effective
complexity provides a way to break the degeneracy between the
quality--of--fit term ($\LL_0$) and the Occam's razor factor in
the marginal likelihood, and enables us to establish whether the
data is good enough to support the introduction of the additional
parameters of model $B$.

To summarize, we are confronted with the following scenarios:
\begin{enumerate}
 \item $p(d|B) \gg p(d|A)$: model $B$ is clearly favored over
model $A$ and the increased number of parameters is justified by
the data.
 \item $p(d|B) \approx p(d|A)$ and $\CC_b(B) > \CC_b(A)$:
the quality of the data is sufficient to measure the additional
parameters of the more complicated model, but they do not improve
its likelihood by much. We should prefer model $A$, with less
parameters.
 \item $p(d|B) \approx p(d|A)$ and $\CC_b(B) \approx \CC_b(A)$:
 both models have a comparable likelihood and the effective
number of parameters is about the same. In this case the data is
not good enough to measure the additional parameters of the more
complicated model and we cannot draw any conclusions as to whether
the additional complexity is warranted.
\end{enumerate}
We illustrate these cases by computing the model likelihood and
effective complexity of a toy model in the next section.

\subsection{An illustrative example\label{sec:toymodel}}

As a specific example
of the linear model described in section \ref{sec:linmod1}, we
consider the classic problem of fitting data drawn from a
polynomial of unknown degree. The models that we test against the
data are a collection of polynomials of increasing order, with
input complexity $\CC_0 = n$, where $n$ is the order of the
polynomial. The question is then whether our model selection can
correctly recover the order $m$ of the polynomial from which the
data are actually drawn.

\begin{figure}[ht]
\begin{center}
\includegraphics[width=70mm]{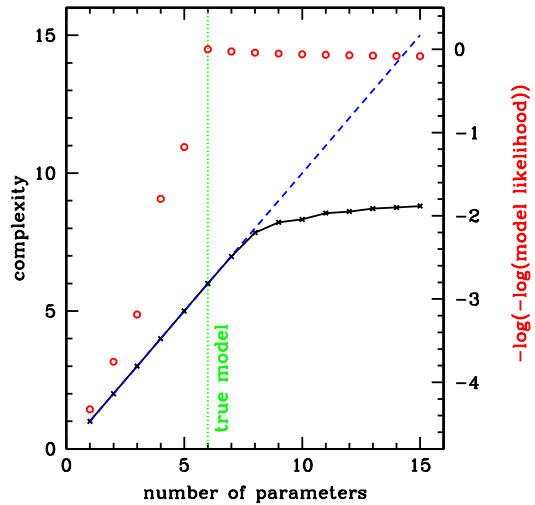} \\
\caption{\label{fig:c1} Bayesian effective complexity $C_b$ (solid
black line, left--hand vertical scale) and model likelihood (red
circles, right--hand scale) as a function of the number of
parameters, for $d=10$ data points with small noise. The dashed
blue line is the number of parameters for reference. The errorbars
on the model likelihood values are smaller than the symbols on
this scale, while the Bayesian complexity is independent of the
noise realization (\ie, error--free) for linear models. The
Bayesian analysis correctly concludes that the best model is the
one with $n=6$ parameters.}
\end{center}
\end{figure}

The data covariance matrix is taken to be diagonal and with a
common standard deviation for all points, $s$, while the prior
over the polynomial coefficients is a multivariate Gaussian with
covariance matrix given by the identity matrix. For
definitiveness, we will take the underlying model from which the
data are generated to have $m=6$ parameters. First we draw $p=10$
data points with noise $s=1/200$. We plot in Figure \ref{fig:c1}
the resulting model likelihood and effective complexity as a
function of the input model complexity, $n$. The likelihood of the
models increases rapidly until $n=m$ and then decreases slowly,
signaling that $n>6$ parameters are not justified. (We plot the
logarithm of the logarithm of the model likelihood as the models
with $n<6$ parameters are highly disfavored and would otherwise
not fit onto the figure -- an example of case (1) of the list in
the previous section). The effective complexity on the other hand
continues to grow until $n\sim p$, at which point the data is
unable to constrain more complex models and $C_b$ becomes
constant. We conclude that the model with $n=6$ is the one
preferred by data and that additional parameters are not needed,
although the data could have supported them. This is case (2) in
the list of the previous section.

\begin{figure}[ht]
\begin{center}
\includegraphics[width=70mm]{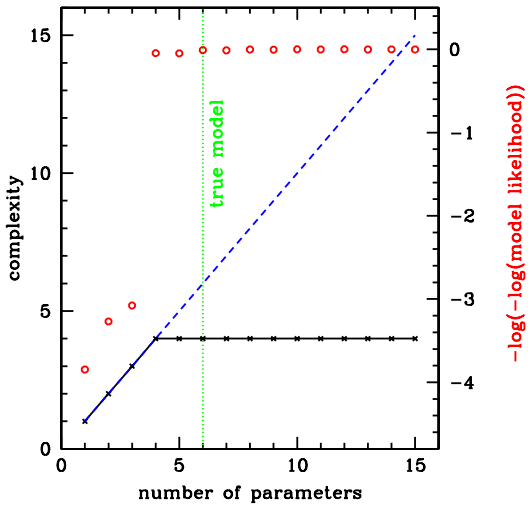} \\
\caption{\label{fig:c2} As in Figure \ref{fig:c1}, but now using
only $p=4$ data points. The maximum effective complexity that the
data can support is $\CC_b = 4$, and the flattening of the model
likelihood at the same value does not allow to conclude that
models with more parameters are disfavored.}
\end{center}
\end{figure}

Next we decrease the number of data points to $p=4$, in which case
we obviously cannot recover more than four parameters. It comes as
no surprise that the model likelihood stops increasing at $n=4$,
see Figure~\ref{fig:c2}. But the effective complexity also
flattens at $n=4$, which means that the data cannot deal with more
than four parameters, irrespective of the underlying model! In
this case, corresponding to point (3) of the list in the previous
section, we conclude that the available data do not support more
than 4 effective parameters. On the other hand, we recognize that
the flattening of the model likelihood at $n=4$ does not
necessarily mean that the underlying model has four parameters. We
thus must hold our judgment until better data become available.

\begin{figure}[ht]
\begin{center}
\includegraphics[width=70mm]{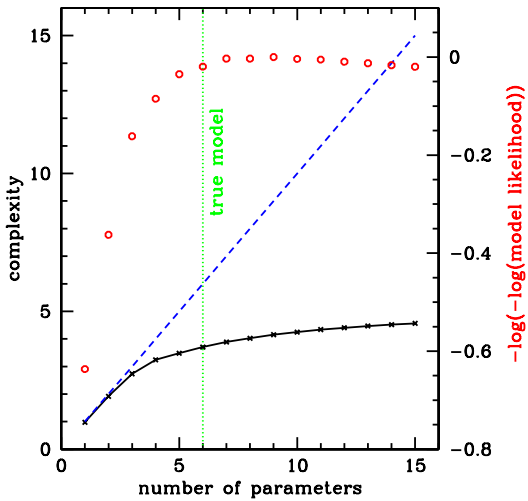} \\
\caption{\label{fig:c3} As in Figure~\ref{fig:c1}, but now with
$p=10$ data points and large noise. As in Figure~\ref{fig:c2}, the
maximum complexity supported by the data is smaller than the
underlying true model complexity, $m=6$.}
\end{center}
\end{figure}

As an alternative to decreasing the number of data points, we can achieve a
similar data degradation effect by keeping $p=10$ data points but
by increasing the noise to $s=1$. We obtain a result similar to
the previous case, which is plotted in
Figure~\ref{fig:c3}.

We conclude by emphasizing once more that in general the outcome
of model selection based on assessing the model likelihood and
effective complexity depends on the interplay of two factors. The
first is the predictive power of the model, as encoded in the
prior. The second is the constraining power of the data.

\section{How many parameters does the CMB need? \label{sec:cosmo}}

We now apply the above tools to the question of how many
cosmological parameters are necessary to describe current cosmic
microwave background (CMB) anisotropy measurements.  We make use
of the following CMB data: WMAP, ACBAR, CBI, VSA and
Boomerang 2003. To provide an additional regularization
(especially in view of including spatial curvature) we also use
the HST limits on the Hubble parameter, $H_0 = 72\pm8$ km/s/Mpc.
We should note that this strongly increases the power of the CMB
data. We use both the first-year WMAP alone (WMAP1) as well as the
data for the first three years (WMAP3).

For each set of cosmological parameters we create a converged MCMC
chain using the publicly available code \texttt{cosmoMC}
\cite{cosmomc}. We then compute the Bayesian complexity from the
chain through Eq.~(\ref{eq:complex_as_chi}). The model likelihood
is evaluated with the Savage-Dickey method (see
\cite{trotta_evidence} and references therein): For a model that
is nested within a larger model by fixing one parameter,
$\theta_f$, to a value $\theta_0$, the Bayes factor between the
two models is given by the posterior of the larger model at
$\theta_f = \theta_0$ (normalized and marginalized over all other
parameters) divided by the prior at this point. In this way it is
possible to derive all the model likelihoods in a hierarchy,
starting from the most complex model (which is assigned an
arbitrary model likelihood, in our case 1). Since errors tend to
accumulate through the intermediate steps necessary to reach the
simpler models, and as a cross--check, we additionally computed
the model likelihoods with nested sampling~\cite{nestsamp} for
the WMAP1 data. Within
the errorbars, we did not find any appreciable discrepancy between
the two methods.

In this analysis we use four basic cosmological parameters, namely
 \be
  b_4 = \{\Omega_b h^2,\Omega_m h^2,\theta_\star,A_s\},
  \ee where
$\Omega_b$ ($\Omega_m$) is the baryon (matter) density relative to
the critical energy density, $h = H_0/100$ km/s/Mpc is the fudge
factor, $\theta_\star$ is the ratio of the sound horizon to the
last scattering angular diameter distance and $A_s = \ln P(k_0)$,
with $P(k_0)$ the power spectrum of adiabatic density fluctuations
at a scale $k_0 = 0.05$ Mpc$^{-1}$. We then add three more
parameters in various combinations, to study whether they are
necessary and supported by the observations. The additional
parameters are: The reionization optical depth $\tau$, the scalar
spectral index $n_s$ and the spatial curvature (parameterized
by its contribution to the Hubble equation, $\Omega_K$).

The scalar spectral index is either
fixed to $n_s=1$ (the case of a scale invariant power spectrum of
initial fluctuations), or else chosen with a Gaussian prior,
$n_s=1\pm0.1$. We find that $\pi(n_s=1)=1/(0.1\sqrt{2\pi})\approx
4$. We choose the prior of the reionization optical depth to
reflect our current lack of understanding of how reionization
proceeds. We choose it flat within $0\leq\tau\leq0.15$ and then
add an exponential falloff:
 \be
 \pi(\tau )\propto\exp\left(-\frac{\tau-0.15}{0.05}\right) \text{ for } \tau > 0.15.
 \ee
The models without this parameter have $\tau=0$, and
$\pi(\tau=0)=5$. The curvature contribution is either set to zero, $\Omega_K=0$, if
we do not include the parameter, or else is used with a flat prior
$-1\leq\Omega_K\leq1$. The value of the prior for a flat universe
is $\pi(\Omega_K=0)=1/2$. We find that adding curvature as a free
parameter when using the WMAP 3yr data leads to a non-Gaussian
posterior for which the mean as a point-estimate for the KL divergence
in Eq.~(\ref{eq:C_b_rewritten}) is not representative. We opt here for a slightly
modified estimator, given by the average of the $\chi^2$ evaluated at the mean
and the mode of the posterior. For a Gaussian posterior this
reduces to the mean point-estimator but it appears to be somewhat
more stable.

\begin{table*}[ht]
\begin{minipage}{150mm}
\begin{tabular}{|l|c|cc|l|}
\hline
Model                    & Model likelihood & $\CC_0$ & Effective complexity  & Comments        \\
\hline
$b_4+n_s+\Omega_K+\tau$  & 1      &7       & $6.9\pm0.3$           & Too many parameters\\
\hline
$b_4+n_s+\Omega_K$       & $0.035 \pm 0.005$ & 6       & $6.3\pm0.2$           & $\Omega_K$ disfavored\\
$b_4+n_s+\tau$           & $48\pm2$     & 6       & $6.0\pm0.04$          & $n_s + \tau$ favored\\
$b_4+\Omega_K+\tau$      & $0.04 \pm 0.01$  & 6    & $6.4\pm0.3$           & $\Omega_K$ disfavored\\
\hline
$b_4+n_s$                & $2.2\pm0.3$  & 5       & $5.0\pm0.04$          & $n_s$ necessary \\
$b_4+\Omega_K$           & $(1.5\pm0.5)\times10^{-5}$ & 5       & $4.8\pm0.04$  & $\Omega_K$ disfavored      \\
$b_4+\tau$               & $3.5\pm1$     & 5       & $4.9\pm0.04$          & $\tau$ necessary \\
\hline
$b_4$                    & $(1.5\pm0.5)\times10^{-3}$ & 4       & $4.0\pm0.04$  & Strongly disfavored        \\
\hline
\end{tabular}
\caption{Relative model likelihood (normalized to the model with
the most parameters) with WMAP 3yr data 
and effective Bayesian complexity for the
models discussed in the text. $\CC_0$ gives the number of
parameters of the model. 
The error on the effective complexity
was computed from random sub--chains and represents only the
statistical error. \label{tab:cmbcplx3}}
\end{minipage}
\end{table*}

\begin{table*}[ht]
\begin{minipage}{150mm}
\begin{tabular}{|l|c|c|}
\hline
Model                    & Model likelihood & Effective complexity  \\
\hline
$b_4+n_s+\Omega_K+\tau$  & 1                & $6.2\pm0.1$           \\
\hline
$b_4+n_s+\Omega_K$       & $0.06 \pm 0.01$  & $6.0\pm0.3$           \\
$b_4+n_s+\tau$           & $42\pm5$         & $5.5\pm0.02$          \\
$b_4+\Omega_K+\tau$      & $0.68 \pm 0.15$  & $5.6\pm0.2$           \\
\hline
$b_4+n_s$                & $4.5\pm0.9$      & $4.9\pm0.03$          \\
$b_4+\Omega_K$           & $(1.5\pm0.5)\times10^{-4}$ & $5.0\pm0.09$  \\
$b_4+\tau$               & $17\pm5$         & $4.8\pm0.03$          \\
\hline
$b_4$                    & $(2.3\pm0.6)\times10^{-3}$ & $4.0\pm0.05$  \\
\hline
\end{tabular}
\caption{Model likelihood and complexity as inferred from
the first year WMAP data, for comparison with the values
in table \ref{tab:cmbcplx3}. We see how the WMAP3 data
increases the model likelihood of $b_4+n_s+\tau$ relative
to the simpler models $b_4+\tau$ and $b_4+n_s$.\label{tab:cmbcplx1}}
\end{minipage}
\end{table*}

\begin{figure}[ht]
\begin{center}
\includegraphics[width=70mm]{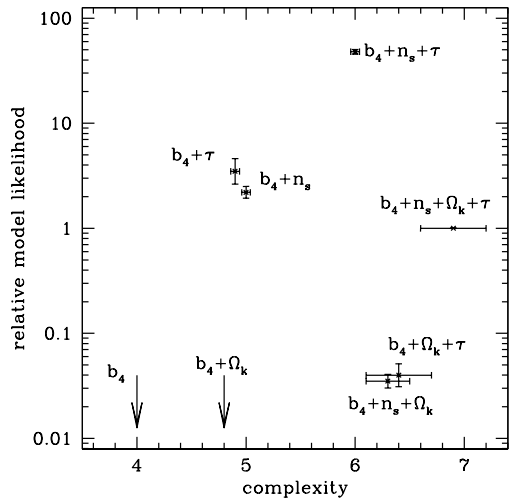} \\
\caption{\label{fig:cmbcomplex} We plot the model likelihood
(normalized to the model with the most parameters) versus the
Bayesian effective complexity for the models of
Table~\ref{tab:cmbcplx3}. A downward--pointing arrow indicates
the Bayesian complexity of models that lie outside the boundary of
the figure. }
\end{center}
\end{figure}

We quote our results in Table~\ref{tab:cmbcplx3} (WMAP3)
and Table~\ref{tab:cmbcplx1} (WMAP1), while
Figure~\ref{fig:cmbcomplex} gives a graphical representation. The
model likelihoods are quoted relative to the model with the most
parameters. We find that using WMAP1 we can measure all
the parameters for the models with
four and five parameters. For $\CC_0 > 5$, the
effective complexity increases more slowly than the number of
input parameters, but we can still measure at least six parameters
with CMB+HST. With WMAP3 we can
measure all six parameters of the $b_4+n_s+\tau$ model.
We conclude that the new WMAP3 data augmented by the HST
determination of $H_0$ can measure all seven parameters considered in this
analysis.

Taking into account the model likelihood values, we find that the
models $b_4$, $b_4+\Omega_K$ and (to a lesser extent)
$b_4+n_s+\Omega_K$ as well as $b_4+\tau+\Omega_K$ are strongly disfavoured. This shows that
$\Omega_K=0$ is preferred by current data, in agreement with the
result of Ref.~\cite{trotta_evidence}. In general, adding in a
non--zero spatial curvature leads to a well measurable decrease in
the model probability that, together with the increase in
effective model complexity, reinforces our belief that $\Omega_K$
can be safely neglected for the time being. Of course this result
is partially a reflection of our choice of prior on $\Omega_K$.
However, it is important to remember that had we halved the range
of this prior, the likelihoods for models with non--zero curvature
would have only doubled. This would not change the results
significantly. Alternatively, an inflation--motivated prior of the
type $|\Omega_K|\ll 1$ would render the parameter unmeasured and
irrelevant. In this case adding it would not change the effective
complexity or the model likelihood at all.

We also find that the basic set $b_4$ must be augmented by either
$n_s$ or $\tau$. The inclusion of both parameters at the same time
was optional with the first year WMAP data only, but using WMAP3 we
find that $b_4 + n_s + \tau$ has a significantly higher model
likelihood than all other models investigated here. Also, where
with WMAP1 we only gained half an effective parameter 
when going from $b_4 + \tau$ or $b_4 + n_s$ to $b_4 + n_s + \tau$,
we now gain one full parameter. Thus we can now measure both
parameters at the same time. 

Overall, we conclude that $b_4+\tau$ was a good and sufficient base
model until a few months ago. Now $b_4 + n_s + \tau$ should be
used. A wider prior on
$n_s$ would have only a minimal impact on the complexity of models
including a tilt, since $n_s$ seems to be rather well--measured
when considered alone (ie, not in combination with $\tau$).
Inclusion of a non--vanishing curvature is discouraged by Bayesian
model comparison. We find that a model with 6 parameters is
sufficient to explain the current CMB data, even though all
seven effective parameters can be constrained now. This analysis
demonstrates that the Bayesian complexity estimator
(\ref{eq:complex_as_chi}) works with real--world data and gives
useful additional information for model comparison.

\section{Conclusions \label{sec:conclusions}}

In this work we introduced the Bayesian complexity as a measure of
the effective number of parameters in a model. We discussed
extensively its properties and its usefulness in the context of
linear models, where it can be computed analytically. We showed
that it corresponds to the number of parameters for which the
width of the posterior probability distribution is significantly
narrower than the width of the prior probability distribution.
These parameters can be considered to have been well measured by
the data given our prior assumptions in the model.

We also showed that for linear models the Bayesian complexity
probes the trace of the posterior covariance matrix, while the
model likelihood is sensitive to the determinant. We argued that
the Bayesian complexity allows to test for cases where the data is
not informative enough for the model likelihood to be a reliable
indicator of model performance, and provided an explicit example.

Finally, we applied the combination of model likelihood and
Bayesian complexity to the question of how many (and which)
parameters are measured by current CMB data, complemented by the
HST limit on the Hubble parameter. We limited ourselves to the family of
$\Lambda$CDM models with a power--law spectrum of primordial
perturbations. We demonstrated that -- in addition to the energy
density in baryons and matter, the CMB peak location parameter
$\theta_\star$ and the amplitude of the initial perturbations --
we need to consider now both the reionisation optical depth
and the scalar spectral index. Non--flat models are disfavoured. The effective
complexity shows that the CMB data can measure all seven
parameters in this scheme.

As the Bayesian complexity is very easy to compute from a MCMC
chain, we hope that it will be used routinely in future data
analyses in conjunction with the model likelihood for model
building assessment. It will help to determine if the data is
informative enough to measure the parameters under consideration.
Further work is needed to study the performance of the Bayesian
complexity in situations with a strongly non--Gaussian posterior.

\begin{acknowledgments}

We thank Andrew Liddle, Pia Mukherjee and Glenn Starkman for
useful discussions. M.K.\ is supported by the Swiss NSF. R.T.\ is
supported by the Royal Astronomical Society through the Sir Norman
Lockyer Fellowship. D.P.\ is supported by PPARC. The calculations
were performed on the ``myri'' cluster of the University of Geneva
and the UK national cosmology supercomputer (COSMOS) in Cambridge.

\end{acknowledgments}

\appendix

\section{Model likelihood and complexity in the linear case}
\label{ap:computed}

Here we compute first the model likelihood \eqref{eq:evidence} for
the linear model \eqref{eq:like}, using the Gaussian prior
\eqref{eq:gaussprior}. An analogous computation can be found in
\cite{dic}. Returning to Bayes theorem (\ref{eq:bayes}), the
posterior pdf is given by a multinormal Gaussian with Fisher
information matrix $F$
    \be F = L + P
\label{eq:post} \ee
 and mean $\bar{\theta}$ given by
 \be
 \bar{\theta} = F^{-1}L \theta_0.
 \ee
The model likelihood \eqref{eq:evidence} evaluates to
  \be \label{eq:modlike}
  \begin{aligned}
    p(d | \mdl) & = \LL_0 \frac{|F|^{-1/2}}{|P|^{-1/2}}
  \exp\left[-\frac{1}{2}\theta_0^t(L - LF^{-1}L )\theta_0
  \right]\\
  & = \LL_0 \frac{|F|^{-1/2}}{|P|^{-1/2}}
  \exp\left[-\frac{1}{2}(\theta_0^t L \theta_0 - \bar{\theta}^tF\bar{\theta})
  \right].
  \end{aligned}
   \ee
This can be easily interpreted by looking at its three components:
the quality of fit of the model is encoded in $\LL_0$, which
represents the best--fit likelihood. Thus a model that fits the
data better will be favored by this term. The term involving the
determinants of $P$ and $F$ is a volume factor (the so called {\em
Occam's factor}). As $|P|\leq|F|$, it penalizes models with a
large volume of wasted parameter space, \ie\ those for which the
parameter space volume $|F|^{-1/2}$ that survives after arrival
of the data is much smaller than the initially available parameter
space under the model prior, $|P|^{-1/2}$. Finally, the
exponential term suppresses the likelihood of models for which the
parameters values that maximize the likelihood, $\theta_0$,
differ appreciably from the expectation value under the posterior,
$\bar{\theta}$. Therefore when we consider a model with an
increased number of parameters we see that its model likelihood
will be larger only if the quality--of--fit increases enough to
offset the penalizing effect of the Occam's factor.

Let us now turn to the computation of the Bayesian complexity,
Eq.~\eqref{eq:complex_as_chi}. Using the posterior mean (denoted
by an overbar) as a point estimator for the effective chi--square,
we obtain from \eqref{eq:defchisq}
 \be \chi^2(\bar{\theta}) = \chi^2(\theta_0) + (\bar{\theta}-\theta_0)^t L
(\bar{\theta}-\theta_0).
 \ee
The expectation value of the $\chi^2$ under the posterior is given
by
 \be
 \overline{\chi^2(\theta)} = \chi^2(\theta_0) + \tr\left( L \langle
(\theta-\theta_0)^t (\theta-\theta_0) \rangle \right) .
 \ee
 We concentrate on the second term and write $(\theta-\theta_0) =
(\theta-\bar{\theta}) + (\bar{\theta} - \theta_0) = u + v$. The
total term in the expectation value then becomes $\langle u^t u +
u^t v + v^t u + v^t v \rangle$ . The first term of this expression
is just the posterior covariance matrix $\langle u^t u \rangle =
F^{-1}$. The last term combines with $\chi^2(\theta_0)$ to
$\chi^2(\bar{\theta})$. The cross terms vanish since $\langle u
\rangle = 0$.

All taken together, we obtain for the complexity
 \bea \CC_b &=& \tr \left( L\langle (\theta-\bar{\theta})^t  (\theta-\bar{\theta}) \rangle\right) \\
&=& \tr  \left( L F^{-1} \right)
 \eea
Using the relation (\ref{eq:post}) we can rewrite the complexity
as \bea
\CC_b &=& \tr\left\{(F-P)F^{-1}\right\} \\
&=& \CC_0 - \tr\left\{P F^{-1}\right\} . \label{eq:dic_gauss}
 \eea
Thus while the model likelihood depends on the determinant of the
Fisher matrices, the complexity depends on their trace. Another
important point worth highlighting is that for linear models the
complexity does not depend on the degree of overlap between the
prior and the posterior, nor on the noise realization (as long as
the noise covariance matrix is known).

\end{document}